\documentclass{PoS}

\title{Solution of the Bethe-Salpeter equation  in Minkowski
space for a  two fermion system}

\ShortTitle{Solution of the Bethe-Salpeter equation  in Minkowski
space for a  two fermion system}

\author{J.~Carbonell$^a$ and \speaker{V.A.~Karmanov}$^{b}$\\
\llap{$^a$}Laboratoire de Physique Subatomique et Cosmologie, CNRS/IN2P3,\\
53 avenue des Martyrs, 38026 Grenoble, France\\
E-mail: \email{carbonel@lpsc.in2p3.fr}\\
\llap{$^b$}Lebedev Physical Institute,\\
Leninsky Prospekt 53, 119991 Moscow, Russia\\
E-mail: \email{karmanov@sci.lebedev.ru}}

\abstract{The  method of solving the Bethe-Salpeter equation in
Minkowski space, developed previously for spinless particles
\cite{lc05,bs1}, is extended to a system of two fermions. The
method is based on the Nakanishi integral representation of the
amplitude and on  projecting the  equation on the light-front
plane. The singularities in the  projected  two-fermion kernel are
regularized  without modifying the original BS amplitudes. The
numerical solutions  for the $J=0$  bound state with the scalar,
pseudoscalar and massless vector  exchange kernels are found.
Binding energies are in close agreement with the Euclidean
results. Corresponding amplitudes in Minkowski space are
obtained.}

\FullConference{Light Cone 2010: Relativistic Hadronic and
Particle
Physics\\
                 June 14-18, 2010\\
                 Valencia, Spain}

\begin{document}

\section{Introduction}

Bethe-Salpeter  (BS) equation  for a relativistic bound system was
initially formulated in the Minkowski space \cite{BS}. It
determines the binding energy and the BS amplitude. However, in
practice, finding the solution in Minkowski space is made
difficult due its singular behaviour. The singularities are
integrable, but the   standard approaches for solving integral
equation fail. To circumvent this problem, the BS equation is
usually transformed, by means of the Wick rotation,  into
Euclidean momentum space.

Some attempts have been recently made to obtain the Minkowski BS
amplitudes. The approach proposed in \cite{KW} is based on the
integral representation of the amplitudes and solutions have been
obtained for the ladder scalar case \cite{KW,SA_PRD67_2003} as
well as, under some simplifying ansatz, for the fermionic one
\cite{sauli}. Another approach \cite{bbpr} relies on a separable
approximation of the kernel which leads to analytic solutions.

In previous works \cite{lc05,bs1} we have proposed a new method to
find the BS amplitude in Minkowski space and applied it to the
system of two spinless  particles. Like in the papers
\cite{KW,SA_PRD67_2003,sauli}, it is based on Nakanishi integral
representation \cite{nakanishi} of the BS amplitude. The main
difference between our approach and  those followed in
\cite{KW,SA_PRD67_2003,sauli} is the fact that we use the
light-front projection. This eliminates the singularities related
to the BS Minko\-w\-s\-ki amplitudes. The method is valid for any
kernel given by the irreducible Feynman graphs.

We present in this contribution a generalisation  to the fermion
systems of our preceding works \cite{lc05,bs1}. A more detailed
version  can be found in \cite{ck_2f}. We will see that the direct
application to the fermionic kernels of the method used in the
spinless case, is however married with some numerical
difficulties. Although they could be overcome by a proper
treatment of the singularities, in this work we propose  an
alternative method allowing to solve the BS equation for two
fermions in Minkowski space with the same degree of accuracy than
for the scalar case. The numerical applications are here limited
to the $J^{\pi}=0^+$ state.

The main steps in deriving system of equations for the Nakanishi
weight functions are explained in sect. \ref{deriv}, starting from
the original BS equation. Numerical results for the scalar,
pseudoscalar and massless vector exchange couplings are presented
in sect. \ref{num}. Section \ref{concl} contains concluding
remarks.

%%%%%%%%%%%%%%%%%%%%%%%%%%%%%%%%%%%%%%%%%%%%%
\section{System of equations}\label{deriv}

We have considered the following fermion ($\Psi,m$) - meson
($\phi,\mu$) interaction Lagrangians:

\noindent ({\it i})~Scalar coupling ${\cal L}_{int} ^{(s)}= g\,
\bar{\Psi} \, \Psi\phi$.

\noindent ({\it ii})~Pseudoscalar coupling ${\cal L}_{int}^{(ps)}=
ig\, \bar{\Psi}\gamma_5  \Psi\phi$.

\noindent ({\it iii})~Massless vector exchange ${\cal L}_{int}
^{(v)}=  \bar{\Psi} \gamma^{\mu}\, \Psi \phi_{\mu}$. with
$\Pi_{\mu\nu}=-i{g_{\mu\nu}/q^2}$  as vector propagator.

Each interaction vertex has been regularized with  a vertex form
factor $F(k-k')$ by  the replacement $ g \to gF(k-k') $ and we
have chosen $F$  in the form:
\begin{equation}\label{ffN}
F(q)=\frac{\mu^2-\Lambda^2}{q^2-\Lambda^2+i\epsilon}.
\end{equation}

Let us first consider the case of the scalar  coupling and  the
corresponding  ladder   kernel. The BS equation for the amplitude
$\Phi$ reads:
\begin{equation} \label{bsf1}
\Phi(k,p)=\frac{i(m+\frac{1}{2}\hat{p}+\hat{k})}
{(\frac{1}{2}p+k)^2-m^2+i\epsilon} \left[\int
\frac{\mbox{d}^4k'}{(2\pi)^4}\;\Phi(k',p)
\frac{(-ig^2)\,F^2(k-k')}{(k-k')^2-\mu^2+i\epsilon}\right]
\frac{i(m-\frac{1}{2}\hat{p}+\hat{k})}{(\frac{1}{2}p-k)^2-m^2+i\epsilon},
\end{equation}
where $p=k_1+k_2$, $k=(k_1-k_2)/2$,  $k'=(k'_1-k'_2)/2$.

In the case of  $J^{\pi}=0^+$ state, the  BS amplitude has the
following  general form:
\begin{equation}\label{bsf2}
\Phi(k,p)=S_1\phi_1+S_2\phi_2+S_3\phi_3+S_4\phi_4
\end{equation}
where $S_{i}$ are independent spin structures ($4\times 4$
matrices)  and $\phi_{i}$ are scalar functions of $k^2$ and
$p\cdot k$.

The choice of $S_i$ is to some extent arbitrary.  To benefit from
useful orthogonality properties we have taken
$$
 S_1= \gamma_5,\quad
 S_2= \frac{1}{M}\hat{p}\,\gamma_5,\quad
 S_3=\frac{k\cdot p}{M^3}\hat{p}\,\gamma_5-\frac{1}{M}\hat{k}\,\gamma_5, \quad
 S_4= \frac{i}{M^2}\sigma_{\mu\nu}p_{\mu}k_{\nu}\,\gamma_5,
$$
where $ \sigma_{\mu\nu}=\frac{i}{2}(\gamma_{\mu}\gamma_{\nu}-
\gamma_{\nu}\gamma_{\mu})$. The antisymmetry of the amplitude
(\ref{bsf2}) with respect to  the permutation $1\leftrightarrow 2$
implies for the scalar functions: $\phi_{1,2,4}(k,p) =
\phi_{1,2,4}(-k,p)$,  $\phi_{3}(k,p) =-\phi_{3}(-k,p)$.

A decomposition similar to (\ref{bsf2}) was used in \cite{sauli}
to solve the BS equation for a quark-antiquark system but the
solution was approximated keeping only the first term $S_1\phi_1$.

We substitute (\ref{bsf2}) in eq. (\ref{bsf1}), multiply it by
$S_{i}$ and take traces. As we will see (left panel in fig.
\ref{V14} below), the kernel in the resulted equation, in contrast
to the spinless case, is still singular. These singularities are
integrable. They do not prevent from finding numerical solution,
but they reduce its precision. This can be avoided by a proper
regularization of equation, multiplying both sides of it by the
factor
\begin{equation}\label{eta}
\eta(k,p)=\frac{(m^2-L^2)}{\left[(\frac{p}{2}+k)^2-L^2+i\epsilon\right]}
\frac{(m^2-L^2)}{\left[(\frac{p}{2}-k)^2-L^2+i\epsilon\right]}
\end{equation}
This factor has the form of a product of two scalar propagators
with mass $L$. It plays the role of form factor suppressing the
high off-mass shell values of the constituent four-momenta
$k^2_{1,2}=(\frac{p}{2}\pm k)^2$ and tends to 1  when $L\to
\infty$. In this way, we get the following system of equations for
the invariant functions $\phi_{i}$:
\begin{eqnarray}\label{bsf4p}
\eta(k,p)\;\phi_i(k,p)&=&\frac{\eta(k,p)}
{[(\frac{p}{2}+k)^2-m^2+i\epsilon]
[(\frac{p}{2}-k)^2-m^2+i\epsilon]} \nonumber\\
&\times& \int \frac{\mbox{d}^4k'}{(2\pi)^4} \frac{i
g^2\,F^2(k-k')}{(k-k')^2-\mu^2+i\epsilon}
\sum_{j=1}^4c_{ij}(k,k',p)\phi_j(k',p),
\end{eqnarray}

Since $\eta(k,p)\neq 0$, the equation thus obtained is strictly
equivalent to the original one. We will see however that, due to
the presence of the $\eta$ factor, the light front projection
modifies the resulting kernels which become less singular
functions. The coefficients $c_{ij}$ are determined  by traces and
are given in \cite{ck_2f}.

Then we represent each of the  BS components $\phi_i(k,p)$ by
means of the Nakanishi integral
\begin{equation}\label{Nakanishi_phi}
\phi_i(k,p)=\int_{-1}^1\mbox{d}z'\int_0^{\infty}\mbox{d}\gamma'
 \frac{g_i(\gamma',z')}{\left[  k^2+p\cdot k\; z'
+\frac{1}{4}M^2-m^2 -  \gamma' + i\epsilon\right]^3}.
\end{equation}
and apply the light-front projection to the  set of coupled
equations for the corresponding weight functions $g_i(\gamma,z)$.
As mentioned in the Introduction, this projection, which is an
essential ingredient of our previous works \cite{lc05,bs1,bs2},
consists in replacing $k\to k+ {\omega\over \omega\cdot p}\beta$
in eq. (\ref{bsf4p})  and integrating over $\beta$ in all the real
domain.

The technical details of the light-front projection are similar to
those  explained in ref. \cite{bs1} for the case of the spinless
particles. We obtain in this way a set of coupled two-dimensional
integral equations:
\begin{equation}\label{eq0f}
\int_0^{\infty}\mbox{d}\gamma'\int_{-1}^1\mbox{d}z' \;
V^g(\gamma,z;\gamma',z')\; g_i(\gamma',z') =
\sum_{j}\int_0^{\infty}\mbox{d}\gamma'\int_{-1}^{1}\mbox{d}z'
\;V^d_{ij}(\gamma,z;\gamma',z') g_j(\gamma',z')
\end{equation}
The kernel $V^g$ and also $V^d_{ij}$ for all types of couplings
and states are given in \cite{ck_2f}. These kernels depend on the
parameter $L$. Closer is $L$ to $m$, smoother is the kernel  and
more stable are the  numerical solutions.  However  the    weight
functions $g_i(\gamma,z)$ as well as  the binding energies
provided by (\ref{eq0f})  are independent of $L$.

The kernel  $V_g$  is finite and it vanishes for $z=\pm 1$. For a
fixed values  of $\gamma,z$ and $\gamma'$, $V_g$ is a continuous
function of  $z'$ with a discontinuous derivative at $z'=z$.

\begin{figure}
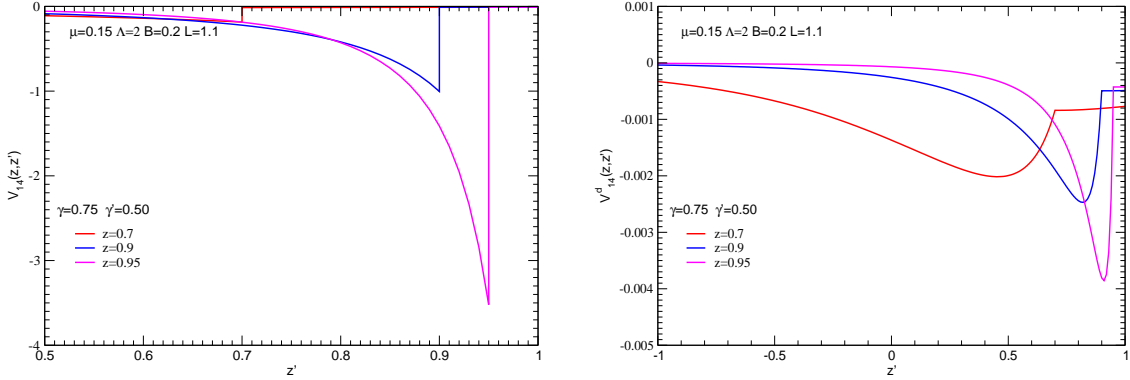
%[ht!]
\vspace{.5cm}
\includegraphics[height=5cm,width=0.47\textwidth]{V14_2.eps}
\hspace{0.5cm}
\includegraphics[height=5cm,width=0.47\textwidth]{V14_d_L_1.1_V2a.eps}
\caption{Left: The kernel matrix elements
$V_{14}(\gamma,z;\gamma',z')$ for $z=0.7,\;0.9,\;0.95$ as a
function of $z'$ and fixed values of $\gamma,\gamma'$. The
discontinuity is finite at a fixed value of $z$ but diverges when
$z\to1$.
\newline
Right: Regularized kernel $V^d_{14}(\gamma,z;\gamma',z')$  v.s.
$z'$ for fixed  values of  $z=0.7,\;0.9,\;0.95$ and  $L=1.1\,m$.}
\label{V14}
\end{figure}

As already mentioned, without $\eta$-factor, most of the kernel
matrix elements $V^d_{ij}$ are singular. Namely, they are
discontinuous at $z'=z$. In some cases -- like {\it e.g.} $V_{14}$
displayed in fig. \ref{V14} (left) -- the value of the
discontinuity, although being finite at fixed value of $z$,
diverges when $z\to\pm1$.

For $\eta\neq 1$, {\it i.e.}, for a finite value of $L$ in
(\ref{eta}), the $z'$-dependence of the regularized kernels is
much more smooth and therefore better adapted  for obtaining
accurate numerical solutions. In fig. \ref{V14} (right) we plotted
the regularized kernel $V^d_{14}$ as a function of $z'$ for the
same arguments $\gamma,z,\gamma'$ and parameters than in  fig.
\ref{V14} (left), where it was calculated without the $\eta(k,p)$
factor. As one can see, the kernel is now a continuous function of
$z'$.

We would like to emphasize again that despite the fact that the
non-regularized and regularized kernels are very different from
each other (compare {\it e.g.} figs. \ref{V14}, left and right)
and that the regularized ones strongly depend on the value of $L$,
they provide -- up to numerical inaccuracies -- the same binding
energies and  weight functions $g_i(\gamma,z)$. We construct in
this way a family of equivalent kernels.

\begin{table}[ht!]
\begin{center}
\caption{Left: Coupling constant $g^2$ as a function of  binding
energy $B$   for  the $J=0$ state with scalar (S), pseudoscalar
(PS) and massless vector (positronium) exchange kernels. The
vertex form factor parameter is $\Lambda=2$ and the parameter of
the $\eta$ factor $L=1.1$
\newline
Right: Coupling constant $g^2_{BS}$ as a function of  binding
energy $B$ for the positronium $J=0$ state in BS equation in the
region of stability without vertex form factor
($\Lambda\to\infty$), {\it i.e.} $g<\pi$. They are compared to the
non relativistic results $g^2_{NR}$. }\label{tab_B_S_Ps}
\vspace{0.3cm}
\begin{tabular}{c|cc |cc |cc |c}
           &  \qquad S                 &                         & \qquad PS  &     & positronium\\ \hline
$\mu$&  0.15           & 0.50        & 0.15      & 0.50 &0.0
\\ \hline
$B$    & $g^2$         & $g^2$     & $g^2$  & $g^2$  & $g^2$              \\
0.01   &   7.813        &  25.23     &   224.8  &  422.3         &     3.265        \\
0.02   &   10.05        &  29.49     &   232.9  &  430.1         &     4.910 \\
0.03   &   11.95        &  33.01     &   238.5  & 435.8          &     6.263 \\
0.04   &   13.69        &  36.19     &   243.1  & 440.4          &    7.457 \\
0.05   &   15.35        &  39.19     &   247.0  &  444.3         &    8.548 \\
0.10   &   23.12        &  52.82     &   262.1  &   459.9        &  13.15  \\
0.20   &   38.32        &  78.25     &   282.9  &   480.7        &   20.43\\
0.30   &   54.20        & 103.8      &   298.6  &   497.4        &  26.50    \\
0.40   &   71.07        & 130.7      &    311.8 &     515.2      &   31.84 \\
0.50  &   86.95         & 157.4      &    323.1 &     525.9      &   36.62 \\
\end{tabular}
\hspace{2cm} % positronium\\
\begin{tabular}{c|cc |cc |cc |c}
%    &$g^2_{NR}$&$g^2_{BS}$               \\ \hline
$B$ &$g^2_{NR}$&$g^2_{BS}$               \\ \hline
0.01   &   2.51          &    3.18        \\
0.02   &   3.55          &    4.65 \\
0.03   &   4.35          &    5.75   \\
0.04   &   5.03          &    6.64    \\
0.05   &   5.62          &    7.38   \\
0.06   &   7.95          &    8.02 \\
0.07   & 11.24          &   8.57 \\
0.08   & 13.77          &   9.06  \\
0.09   & 15.90          &   9.49\\
\end{tabular}
\end{center}
\end{table}

%%%%%%%%%%%%%%%%%%%%%%%%%%%%%%%%%%%%%%%%%%%%%%%%%%%%%
\section{Numerical results}\label{num}

The solutions of eq. (\ref{eq0f}) have been obtained using the
same  techniques than in ref \cite{bs1}. We have computed the
binding energies,  defined as  $B=2m-M$, and BS   amplitudes, for
the $J=0^+$ two fermion system interacting with  massive  scalar
(S) and pseudoscalar (PS) exchange kernels and for the
fermion-antifermion system interacting with massless vector
exchange in Feynman gauge. In the limit of an infinite vertex form
factor parameter $\Lambda\to\infty$, the later case would
correspond to positronium with an arbitrary  value of the coupling
constant. All the results presented in this section are given in
the constituent mass units ($m=1$)   and with $L=1.1$.

\begin{figure}
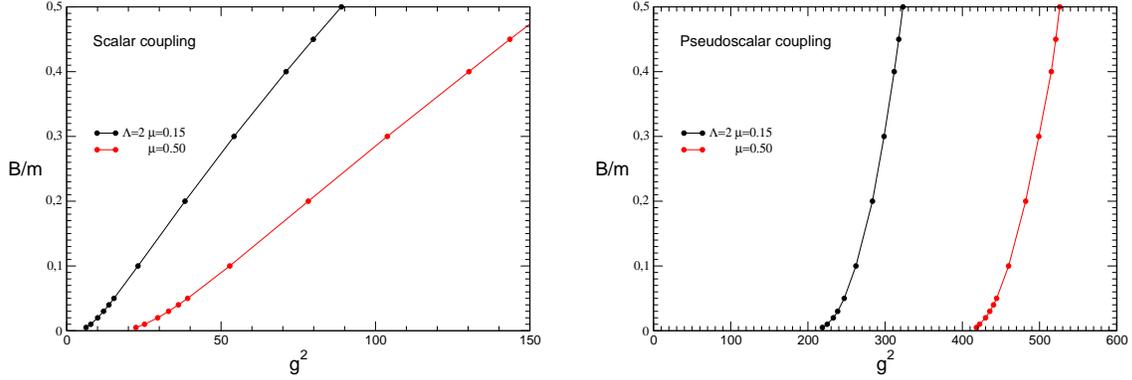
%[ht!]
\vspace{.5cm}
\includegraphics[height=5cm,width=0.47\textwidth]{B_g2_S.eps}
\hspace{0.5cm}
\includegraphics[height=5cm,width=0.47\textwidth]{B_g2_Ps.eps}
\caption{Left: Binding energy for scalar exchange v.s. $g^2$ for
$\Lambda=2$,  $L=1.1$, $\mu=0.15$ and $\mu=0.5$.
\newline
Right: Binding energy for pseudoscalar exchange v.s. $g^2$ for
$\Lambda=2$, $L=1.1$, $\mu=0.15$ and $\mu=0.5$.}\label{Fig_B_g2_S}
\end{figure}

For the scalar and pseudoscalar cases, the binding energies
obtained with the form factor parameter $\Lambda=2$ are given in
the left table \ref{tab_B_S_Ps}. We present the results for
$\mu=0.15$ and $\mu=0.50$  boson masses. They have been compared
to those obtained in a previous calculation in Euclidean space
\cite{dorkin} using a slightly different form factor. Once taken
into account this correction, our scalar  results are in  full
agreement (four digits) with \cite{dorkin}. The pseudoscalar ones
show small discrepancies ($\approx 0.5\%$). We have also computed
the binding energies by directly solving the fermion BS equation
the Euclidean space using a method independent of the one used in
\cite{dorkin}. Our Euclidean results are in full agreement with
those given in the table \ref{tab_B_S_Ps}.

The $B(g^2)$ dependence for the scalar and pseudoscalar couplings
is plotted in figs. \ref{Fig_B_g2_S}. Notice the different $g^2$
scales of both dependences. The pseudoscalar binding energies are
fast increasing functions of $g^2$ and thus more sensitive to the
accuracy of numerical methods. This sharp behaviour was also
exhibit when solving the corresponding light-front  equation
\cite{MCK_PRC68_2003}.

In the positronium case, we found  the existence  a critical value
of the coupling constant $g_c=\pi$. For $g<g_c$, solutions with
finite binding energy exist without form factor ({\it i.e.}, at
$\Lambda\to\infty$).

The ground state  positronium  binding energies without vertex
form factor are given in table  \ref{tab_B_S_Ps} (right)  for
values of the coupling below $g_c$, nonrelativistic results
$g^2_{NR}=8\pi \sqrt{B/m} $ are included for comparison. One can
see that  the relativistic effects  are  repulsive.

\begin{figure}%[ht!]
\vspace{.5cm}
\begin{center}
\includegraphics[height=5cm,width=0.47\textwidth]{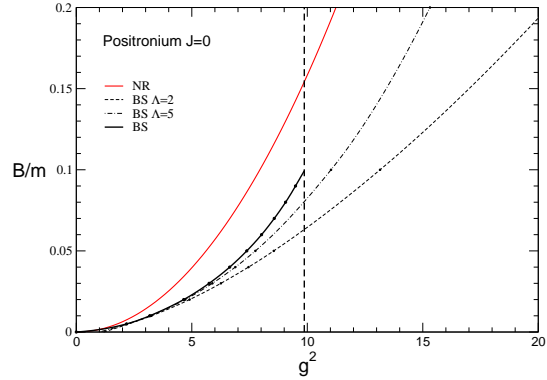}
\end{center}
\caption{Binding energy for J=0 positronium state versus $g^2$
(black solid line) in the stability region $g<g_c=\pi$. Dashed and
dotted-dashed curves correspond to the results for increasing
values of the vertex form factor parameter $\Lambda$. They are
compared to the non relativistic results (red solid line).}
\label{Fig_B_g2_Positronium}
\end{figure}

These results  are displayed in fig. \ref{Fig_B_g2_Positronium}
(black solid line), and compared to the binding energies obtained
with two values of the form factor parameter $\Lambda=2$ (dashed)
and $\Lambda=5$ (dot-dashed). The stability region is limited by a
vertical dotted line  at $g=g_c=\pi$. Beyond this value the
binding energy without form factor becomes infinite and we have
found $B(g\to g_c)\approx0.10$.
The inclusion of the form form factor has a  repulsive effect,
{\it i.e.}  for a fixed value of the coupling constant it provides
a binding energy of the system which is smaller than in the
$\Lambda\to\infty$ limit (no cut-off).

\begin{figure}[ht!]
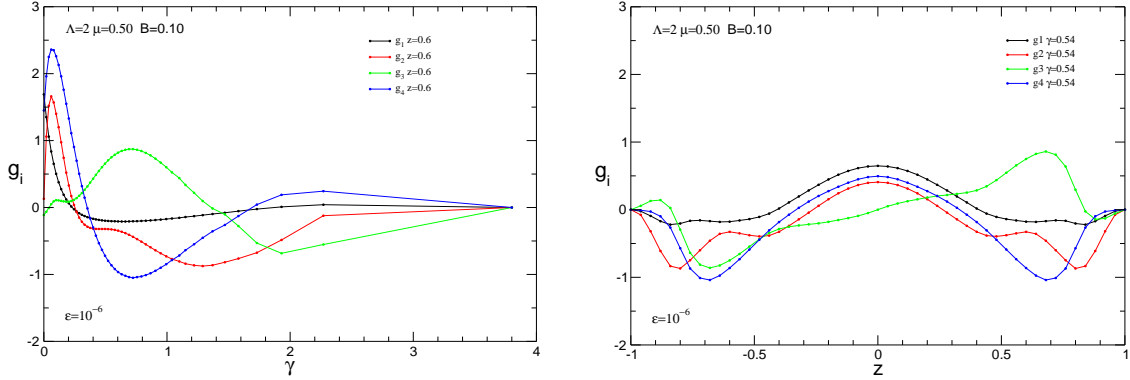

\vspace{.6cm}
\includegraphics[height=5cm,width=0.47\textwidth]{gik_2_0.50_0.10_10M6.eps}
\hspace{0.5cm}
\includegraphics[height=5cm,width=0.47\textwidth]{giz_2_0.50_0.10_10M6.eps}
\caption{Left: Nakanishi weight functions v.s. $\gamma$ for
$z=0.6$, for scalar exchange for $\Lambda=2$,  $L=1.1$, $\mu=0.15$
and $\mu=0.5$. Right:  Nakanishi weight functions v.s. $z$ for
$\gamma=0.54$.\label{gig}}
\end{figure}

Finally, in figs. \ref{gig} we present some examples of the
Nakanishi weigh functions $g_i(\gamma,z)$. They correspond to a
$B=0.1$ state with the scalar coupling  and the same  parameters
$\Lambda=2$, $\mu=0.50$ than in table \ref{tab_B_S_Ps}.  In the
left figure is shown the $\gamma$-dependence for a fixed value of
$z$ and in the right figure  -- the $z$-dependence for a fixed
$\gamma$. Notice the regular behaviour of these functions as well
as their well defined parity with respect to $z$ -- $g_{1,2,4}$
are even and $g_3$ is odd.

\begin{figure}[ht!]
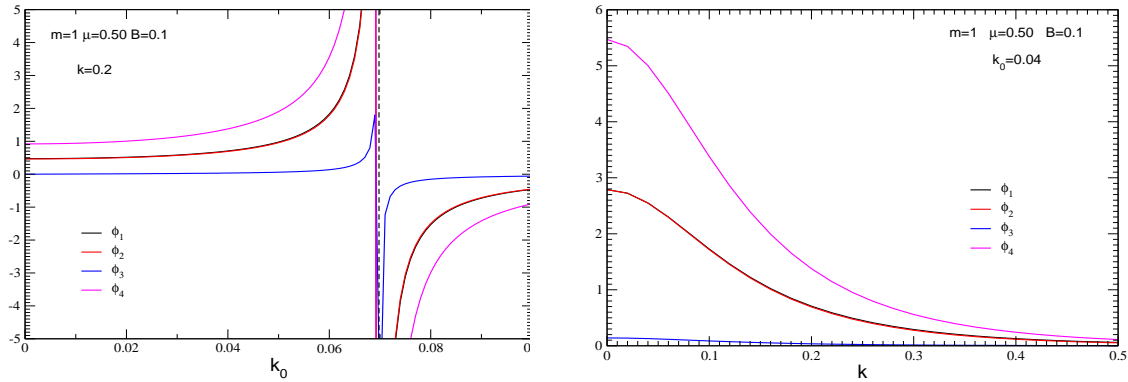

\includegraphics[height=5cm,width=0.47\textwidth]{Phik0_k_0.2_S_10M2.eps}
\hspace{.5cm}
\includegraphics[height=5cm,width=0.47\textwidth]{Phik_k0_0.04_S_10M2.eps}
\caption{Left: Bethe-Salpeter Minkowski amplitudes, corresponding
to fig. 4, v.s. $k_0$ for $k=|\vec{k}|=0.2$. The amplitudes
$\phi_1$ and $\phi_2$ are indistinguishable. Right: The same as at
left v.s. $k=\mid\vec{k}\mid$ for $k_0$=0.04.} \label{Plot_Phi_k0}
\end{figure}

Corresponding BS amplitudes $\phi_i$ are displayed in figs.
\ref{Plot_Phi_k0}. The left panel represents the $k_0$ dependence
of $\phi_i$ for a fixed value of $\mid\vec{k}\mid=0.2$. They
exhibit a singular behaviour which corresponds to the pole of free
propagators $k_0=\epsilon_k- \frac{M}{2}$ in r.-h.-side of eq.
\ref{bsf1}. The right panel  represents the $\mid\vec{k}\mid$
dependence of the amplitudes $\phi_i$ for a fixed value
$k_0=0.04$. For this choice of arguments, the amplitudes are
smooth functions of $\mid\vec{k}\mid$, though they will be also
singular for $k_0> \frac{B}{2}=0.05$.

\vspace{.8cm}

%%%%%%%%%%%%%%%%%%%%%%%%%%%%%%%%%%%%%%%%%%%%
\section{Conclusions}\label{concl}

We have presented a new method to obtain the solutions of the
Bethe-Salpeter equation in  Minkowski space for the two-fermion
system. It  is based on a Nakanishi integral representation of the
amplitude and light-front projection and constitutes a natural
extension our previous works  for the scalar case \cite{lc05,bs1}.

A straightforward generalization of this  approach, however
results into a singular fermionic kernel. In order to smooth this
singularities, a proper regularization of the kernels has been
proposed. This generates a family of strictly equivalent equations
depending on one parameter $L$. Their solution gives the same
binding energies and Nakanishi weight functions $g_i(\gamma,z)$ .

The binding energies for the  scalar and pseudoscalar exchange
kernels  and for massless vector exchange (positronium)  have been
found. They coincide with the ones obtained in Euclidean space,
thus providing a validity test of our method. The solutions for
the scalar and positronium states without vertex form  factor
($\Lambda\to\infty$) are found to be stable below the following
critical values $g_c$ of the coupling constant: $g_c=2\pi$
(scalar) and $g_c=\pi$ (positronium).

The BS amplitudes in Minkowski are obtained in terms of the
computed Nakanishi weight functions. They exhibit a singular
behaviour due to the poles of the free propagators.

%%%%%%%%%%%%%%%%%%%%%%%%%%%%%%%%%%%%%%%%


\begin{thebibliography}{10}
\bibitem{lc05}
V.A. Karmanov, J. Carbonell, {\it Bethe-Salpeter equation in
Minkowski space with cross-ladder kernel}, in proceedings of {\it
The Workshop on Light-Cone QCD and Nonperturbative Hadron Physics,
Cairns, Australia, July 7-15, 2005, Nucl. Phys. B} (Proc.Suppl.)
{\bf 161} (2006) 123[nucl-th/0510051].

\bibitem{bs1}
V.A.~Karmanov and J.~Carbonell, {\it Solving Bethe-Salpeter
equation in Minkowski space, Eur. Phys. J. A} {\bf 27} (2006) 1
[hep-th/0505261].

\bibitem{BS}
E.E.~Salpeter and H.A.~Bethe, {\em A Relativistic Equation for
Bound-State Problems, Phys. Rev.} {\bf 84} (1951) 1232.

\bibitem{KW} K.~Kusaka, A.G.~Williams, {\em Solving the Bethe-Salpeter equation
for scalar theories in Minkowski space, Phys. Rev. D} {\bf 51},
 (1995) 7026.

\bibitem{SA_PRD67_2003} V.~Sauli, J.~Adam, Jr.,
{\em Study of relativistic bound states for scalar theories in
Bethe-Salpeter and Dyson-Schwinger formalism, Phys. Rev. D} {\bf
67} (2003) 085007.

\bibitem{sauli}  V.~Sauli, {\em Solving the Bethe-Salpeter equation for a
pseudoscalar meson in Minkowski space,
 J. Phys. G} {\bf 35} (2008) 035005[arXiv:0802.2955(hep-ph)].

\bibitem{bbpr}
S.G.~Bondarenko, V.V.~Burov, W.-Y.~Pauchy Hwang, E.P.~Rogochaya,
{\em Relativistic multirank interaction kernels of the
neutron-proton system, Nucl. Phys. A} {\bf 832} (2010)
233[arXiv:0810.4470(nucl-th)]; [arXiv:1002.0487(nucl-th)].

\bibitem{nakanishi}
N.~Nakanishi, {\em Partial-Wave Bethe-Salpeter Equation, Phys.
Rev.} {\bf 130} (1963) 1230;\\
{\it Graph Theory and Feynman Integrals}, Gordon and Breach, New
York, 1971.

\bibitem{ck_2f}
J.~Carbonell and V.A.~Karmanov, {\it Solving Bethe-Salpeter
equation for two fermions in Minkowski space}, to appear in {\em
Eur. Phys. J. A.}

\bibitem{bs2}
J.~Carbonell and V.A.~Karmanov, {\em Cross-ladder effects in
Bethe-Salpeter and Light-Front equations, Eur. Phys. J. A} {\bf
27} (2006) 11[hep-th/0505262].

\bibitem{dorkin} S.M.~Dorkin, M.~Beyer, S.S.~Semykh and L.P.~Kaptari,
{\em Two-Fermion Bound States within the Bethe-Salpeter Approach,
Few-Body Syst.} {\bf 42} (2008) 1[arXiv:0708.2146(nucl-th)].

\bibitem{MCK_PRC68_2003} M. Mangin-Brinet, J. Carbonell, V. Karmanov,
{\em Two-fermion relativistic bound states in Light-Front
Dynamics, Phys. Rev. C} {\bf 68}, (2003) 055203[hep-th/0102068].

\end{thebibliography}
\end{document}